\documentclass[aps,prb,superscriptaddress,twocolumn,tightenlines]{revtex4-1}
\usepackage{amssymb}
\usepackage{graphicx}
\usepackage{natbib}
\usepackage{epstopdf}
\usepackage{dcolumn}
\usepackage{bm}
\usepackage{makeidx}
\usepackage{enumerate}
\usepackage {graphicx}
\usepackage {amsmath}
\usepackage {amsfonts}
\usepackage {amssymb}
\usepackage {mathrsfs}
\usepackage {bbm}
\usepackage{subfigure}
\begin{document}
\title{Complex state induced by impurities in multiband superconductors}
\author{Valentin Stanev}
\affiliation{Condensed Matter Theory Center, Department of Physics,
University of Maryland, College Park, MD 20742, USA}
\affiliation{Materials Science Division, Argonne National Laboratory, Argonne,
Illinois 60439, USA}

\author{Alexei E. Koshelev}
\affiliation{Materials Science Division, Argonne National Laboratory, Argonne,
Illinois 60439, USA}
\date{\today }

\begin{abstract}
We study the role of impurities in a two-band superconductor, and elucidate the
nature of the recently predicted transition from $s_{\pm}$ state to $s_{++}$ state
induced by interband impurity scattering. Using a Ginzburg-Landau theory, derived
from microscopic equations, we demonstrate that close to $T_c$ this transition is
necessarily a direct one, but deeper in the superconducting state an intermediate
complex state appears. This state has a distinct order parameter, which breaks the
time-reversal symmetry, and is separated from the $s_{\pm}$ and $s_{++}$ states by
continuous phase transitions. Based on our results, we suggest a phase diagram for
systems with weak repulsive interband pairing, and discuss its relevance to
iron-based superconductors.
\end{abstract}
\maketitle

It has been long recognized that nonmagnetic impurities
strongly influence properties of multiband superconductors\cite{SungJPCS67,Muzikar,
GM1, Kulic, Mishonov, Gurevich}, especially in the case of an order parameter with
sign change between different bands  ($s_{\pm}$ state)\cite{Muzikar, Senga, Bang,
Vorontsov}.
Recently, it has been pointed out that impurities-induced interband scattering can
continuously change the order parameter of a two-band superconductor from $s_{\pm}$
to $s_{\mathrm{++}}$ state\cite{Golubov, Golubov2, GM2} . This is particularly
relevant for iron-based superconductors\cite{LaOFeAs, Paglione}, most of which are
believed to be in some form of the $s_{\pm}$ state, see recent
reviews\cite{Reviews1,Reviews2}.

\begin{figure}[h]
\begin{center}$
\begin{array}{cc}
\includegraphics[width=0.5\textwidth]{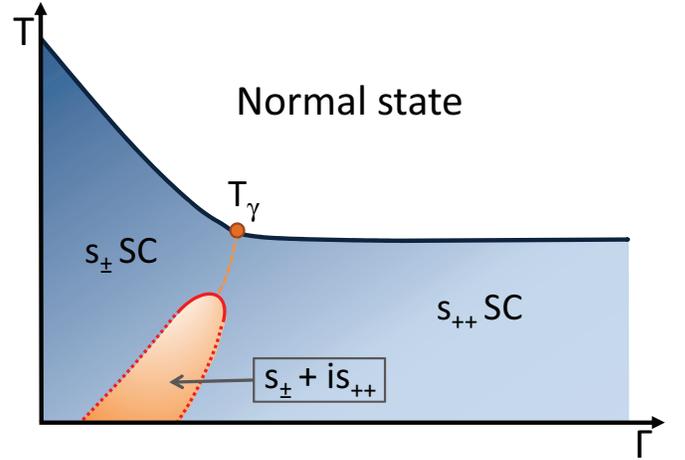}
\end{array}$
\end{center}
\caption{Phase diagram of systems with weak repulsive interband pairing. The $x$-axis represents the interband impurity scattering rate. The orange dashed line denotes the direct $s_{\pm}$ to $s_{\mathrm{++}}$ transition, and the orange region represents the complex $s_{\pm}+ i s_{\mathrm{++}}$ state. The phase transition lines between the complex state and the other states are shown with red, and the dashed red indicate the conjectured extension of the complex state at low temperatures.}
\label{Fig1}
\end{figure}

As we demonstrate in this Letter,
the $s_{\pm}$-to-$s_{\mathrm{++}}$ transformation may follow a nontrivial scenario,
and occur via an intermediate complex state at which a finite phase shift develops
between the gap parameters in the two bands. We derive the simplest possible
two-band Ginzburg-Landau (GL) free energy of the system from microscopic theory, and
show that the presence of interband impurity scattering has important
consequences for the different possible order parameters the theory can support. In
the case of repulsive interband pairing we indeed observe the $s_{\pm}$ to
$s_{\mathrm{++}}$ transition\cite{crossover} with increasing the degree of disorder.
We demonstrate that the transition is necessarily a direct one only close to the
critical line; deeper in the superconducting state the $s_{\pm}$ state gives way to
an intrinsically complex order parameter (which can be thought as an $s_{\pm} + i
s_{\mathrm{++}}$ state), and only then to a pure  $s_{\mathrm{++}}$ state. This
complex state breaks time-reversal symmetry and is separated from the other two
superconducting states by continuous phase transitions. We discuss the reason and
conditions for the appearance of this state. Based on our results, we propose the
phase diagram shown in Fig. \ref{Fig1} for two-band superconductors with weak
repulsive interband coupling.


We consider a system of two parabolic bands, with partial and
total densities of states (DOS) $N_1$, $N_2$, and $N=N_1+N_2$ respectively. The
pairing interactions are described by $2\times 2$ coupling matrix $\hat{\lambda}$,
with $\det [\hat{\lambda}]\equiv w=\lambda_{11} \lambda_{22}-
\lambda_{12}\lambda_{21}$. In the superconducting state there are two gap parameters
$\Delta_1$ and $\Delta_2$, which are assumed to be complex constants for each band
$\Delta_m = |\Delta_m | e^{i\phi_m}$. The relative phase $\varphi=\phi_1- \phi_2$ is
a gauge-invariant quantity, and it is $0$ or $\pi$ in the $s_{\mathrm{++}}$ or
$s_{\pm}$ states respectively. The presence of impurities introduces scattering
rates parametrized by $\gamma_{mn}$, where $m, n=(1,2)$ are the band indices. For
the interband terms ($m \neq n$) we can write $\gamma_{mn}=N_n \Gamma$, with
$\Gamma=n_{imp}\pi u^2$, where $n_{imp}$ and $u$ are the impurities' concentration
and  potential respectively.
On general grounds, point defects, such as atomic substitutions or vacancies, can
scatter carriers with large momentum change and therefore are expected to
give comparable intraband and interband scattering rates. In the case of the
iron-based superconductors this was indeed confirmed by the first-principles
calculations \cite{Kemper}.

Close to the critical temperature the free energy can be expanded in powers of
$|\Delta_1|$ and $|\Delta_2|$. (Although GL theory has been generalized to the case
of multicomponent order parameters without impurities\cite{Tilley, ZD}, the proper
justification of this multiband extension is a matter of ongoing debate\cite{AEK,
Babaev1, Kogan1, Shanenko,GLnote}.) In the presence of impurities  this can be done
systematically, starting from the Usadel equations\citep{Gurevich, AEK}. The
resulting GL free energy up to quartic in $\Delta$ terms can be written as
\begin{eqnarray}
\mathcal{F}_{GL}= \mathcal{F}_{11} + \mathcal{F}_{22} + \mathcal{F}_{12}+ \mathcal{F}_{EM}.
\label{GL1}
\end{eqnarray}
We present the derivation of $\mathcal{F}_{GL}$ from the microscopic theory, and
give exact expressions for its coefficients in the Supplemental
Material\citep{SuplMat}. If the gap parameters are uniform in space and constant
within each band, the intraband impurity scattering rate $\gamma_{mm}$ drops out of
the theory  completely, as a direct consequence of the Anderson
theorem\cite{Anderson}. In contrast, the interband terms play an important role.
The first two terms look similar to the standard GL theory
\begin{eqnarray}
 \mathcal{F}_{mm}(\Delta_i) = a_{mm}|\Delta_{m}|^2 + \frac{b_{mm}}{2}|\Delta_{m}|^4, 
\label{GL2}
\end{eqnarray}
but with $a_{mm}$ and $b_{mm}$ modified by the presence of impurities\cite{SuplMat}.
$\mathcal{F}_{EM}$ combines the electromagnetic field contribution, and the
derivative terms that couple $\Delta_1$ and $\Delta_2$ to the electromagnetic
vector-potential. For the rest of this paper we assume no field and uniform order
parameter, so $\mathcal{F}_{EM}\!=\!0$. The third term in $\mathcal{F}_{GL}$ couples
$\Delta_1$ and $\Delta_2$, and without impurities it is $ 2
a_{12}|\Delta_{1}||\Delta_{2}| \cos{\varphi} $. In the presence of interband scattering
processes, however, $\mathcal{F}_{12}$ becomes more complicated:
\begin{eqnarray}
 \mathcal{F}_{12} = & & 2 a_{12}|\Delta_{1}||\Delta_{2}| \cos{\varphi} +b_{12} |\Delta_{1}|^2|\Delta_{2}|^2 \nonumber\\
+ & & 2(c_{11} |\Delta_{1}|^3|\Delta_{2}| + c_{22} |\Delta_{1}||\Delta_{2}|^3)\cos{\varphi} \nonumber\\
+ & & c_{12} |\Delta_{1}|^2|\Delta_{2}|^2\cos{2\varphi}.
\label{GL3}
\end{eqnarray}
We can see that the presence of impurities
introduces several new quartic interband terms in the GL theory\citep{Ng}.
In the limit $\Gamma\rightarrow 0$ $a_{12}$ becomes proportional to
$\lambda_{12}$ and all other coefficients in Eq.\ (\ref{GL3}) vanish. As a
consequence, for a clean system the only possible solutions for $\varphi$ are $0$
and $\pi$, and which one minimizes $\mathcal{F}_{GL}$ is determined by the sign of
$\lambda_{12}$.
When impurities are
present, this is not necessarily true any more, and other solutions are possible,
due to the $\cos{2\varphi}$ term -- it can destabilize the $s_{\pm}$ and $s_{++}$
states, provided $c_{12}$ is positive\cite{Tanaka1}. Thus, the dirty two-band
superconductor can have quite rich phase diagram.

The critical temperature at a given disorder strength is
determined by the quadratic terms in Eq. (\ref{GL1}). The equation for $T_c$ derived
in the Supplemental Material\cite{SuplMat}
takes the form $\mathrm{det}\left[\mathbb{M}-\mathbb{I}\right]=0$, with $\mathbb{I}$
being the $2 \times 2$ identity matrix, and
\begin{eqnarray}
\mathbb{M}\equiv
\begin{bmatrix}
       \lambda_{11} I_2 \!+\! \lambda_{1} n_1(I_1\!-\! I_2) & \  \lambda_{12} I_2\! +\! \lambda_{1} n_2(I_1\!-\!I_2)
       \\[0.3em]
      \lambda_{21} I_2 \!+\! \lambda_{2} n_1(I_1\!-\! I_2) & \ \lambda_{22} I_2 \! +\! \lambda_{2} n_2(I_1\! -\! I_2)
      \\[0.3em]
     \end{bmatrix}\nonumber.
\end{eqnarray}
We have defined $n_{m}=N_m/N$, $\lambda_m=\lambda_{mm}+\lambda_{mn}$, and
\begin{eqnarray}
I_1= 2 \pi T \sum_{\omega_n>0}^{\omega_0}\frac{1}{|\omega_n|},\ \ I_2= 2 \pi T \sum_{\omega_n>0}^{\omega_0}\frac{1}{|\omega_n|+ \gamma_{12} + \gamma_{21}},\nonumber
\end{eqnarray}
where $\omega_0$ is a high-energy cut-off (e.g., the Debye frequency). In the clean
limit, $\Gamma=0$, this equation gives transition temperature $T_{c0}\approx
1.13\omega_0\exp(-1/\lambda)$, where $\lambda$ is the largest eigenvalue of the
$\hat{\lambda}$-matrix. Note that the interband impurity scattering processes are
always pair-breaking (unless $\Delta_1=\Delta_2$), and suppress $T_c$, in contrast
with the intraband scattering, which has disappeared.

\begin{figure}[h]
\begin{center}$
\begin{array}{cc}
\includegraphics[width=0.48\textwidth]{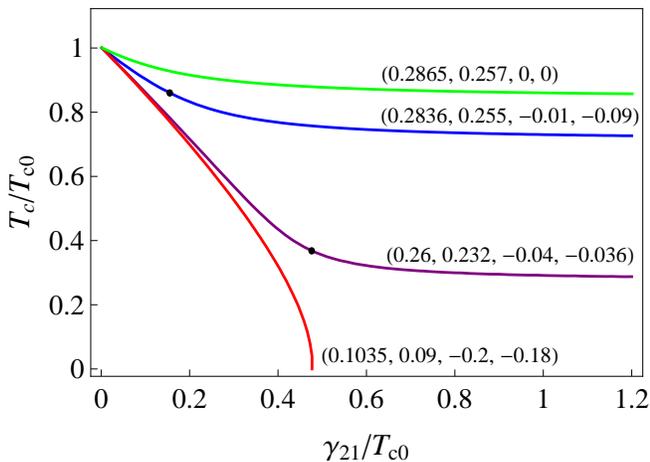}
\end{array}$
\end{center}
\caption{The $T_c$ lines for systems with different $\hat{\lambda}$, as functions of
$\gamma_{21}$. The coupling constants are shown inside the figure, in
$(\lambda_{11}, \lambda_{22}, \lambda_{12}, \lambda_{21})$ format. In the cases of
weak interband pairing (green, blue and purple lines) $T_c$ is initially suppressed,
but eventually saturates. For repulsive \emph{and} strong interband pairing (red
line), superconductivity is completely suppressed by impurities. The dots indicate
the position of the $T_{\gamma}$ points for the blue and the purple curves.}
\label{Fig2}
\end{figure}

In general, the dependence $T_c(\gamma_{mn})$ has to be found numerically but the
extreme dirty limit can be analyzed analytically. Depending on $\hat{\lambda}$,
there are two qualitatively different regimes. If  interband pairing is attractive,
or negative but weak (i.e., when $w$ is positive) no amount of disorder can
completely suppress the superconductivity. In this case the critical temperature in
the extreme dirty limit can be obtained\cite{SuplMat}:
\begin{eqnarray}
T_{c \infty}\!\approx\! 1.13\omega_0\exp\left(\!- \frac{n_1(\lambda_{22}\!-\!\lambda_{12})\!+\! n_2(\lambda_{11}\!-\!\lambda_{21})}{w}\right)\!.
\label{T_cextrim}
\end{eqnarray}
However, if the interband pairing is repulsive \emph{and} strong, such that $w$ is
negative, there is a critical amount of disorder which brings $T_c$ down to zero, in
analogy with the Abrikosov-Gor'kov theory\cite{AG}. Numerical calculation of $T_c$
for the different regimes are shown in Fig. \ref{Fig2}. We see that for some
systems, after the initial drop in $T_c$ from its clean limit $T_{c0}$, the critical
temperature saturates and stays finite in the limit $\Gamma \rightarrow \infty$. The
reason is that the impurity scattering gradually averages the two gaps, and the
closer they get to each other, the less effective the pair-breaking from the
impurities is; thus the superconductivity can survive even in the extremely dirty
regime (in that limit $\Delta_1=\Delta_2$).  The second regime is also easy to
understand --  if the sign change between the gaps is necessary for the existence of
superconductivity (i.e., if the repulsive interband pairing interactions dominate)
then the averaging produced by impurities completely suppresses the order parameter.
Note that although our results are broadly consistent with the ones obtained in Ref.
\onlinecite{Golubov}, our Eq. (\ref{T_cextrim}) somewhat disagrees with the dirty
limit $T_{c}$ derived there, since in our expression the effective coupling constant
is $\langle \lambda^{-1}\rangle^{-1}$ rather than $\langle \lambda \rangle.$

For the rest of this Letter we concentrate on systems with positive $w$ and
repulsive interband pairing  -- as we will see, these are the systems with the most
interesting phase diagram.  We turn to the coefficient $a_{12}$ of the
Josephson-like term $|\Delta_{1}||\Delta_{2}| \cos{\varphi}$, and its evolution with
$\Gamma$. The role of $a_{12}$ is to couple the gaps, guaranteeing that they appear
simultaneously, and close to $T_c$ its sign fixes the relative phase of $\Delta_1$
and $\Delta_2$. In the presence of impurity scattering it is
\begin{eqnarray}
a_{12}= - g - n_1 n_2 N(I_1-I_2),
\label{a12}
\end{eqnarray}
with $g=\lambda_{12}N_1/w=\lambda_{21}N_2/w$. In the clean limit $I_2\rightarrow
I_1$, $a_{12} \rightarrow - g$, and, as a result, $\varphi$ is temperature
independent,
and can only be $0$ or $\pi$. For finite $\Gamma$, however, $a_{12}$ becomes
function of both disorder strength \emph{and} temperature, and can even change its
sign. This has important consequences for the order parameter. Negative $g$ leads to
the $s_{\pm}$ state in the clean limit. However, the second term in Eq. (\ref{a12})
is negative, and for strong disorder it can overcome the $-g$ term. If $T_c$ is not
completely suppressed (i.e., if the intraband pairing dominates), this sign change
of $a_{12}$ means a transition from $s_{\pm}$ to $s_{\mathrm{++}}$ state at the
$T_c(\Gamma)$ line \cite{Golubov}.
This happens at temperature $T_{\gamma}\approx 1.13\omega_0\exp\left[
-(\lambda_{22}-\lambda_{12})/w\right] $\cite{SuplMat}. At this point the bands are
effectively decoupled, and one of them stays normal. At smaller disorder strength
the system condenses in the $s_{\pm}$ state, while at larger disorder strength it
goes into the $s_{\mathrm{++}}$ state.

\begin{figure}[h]
\begin{center}$
\begin{array}{cc}
\includegraphics[width=0.4\textwidth]{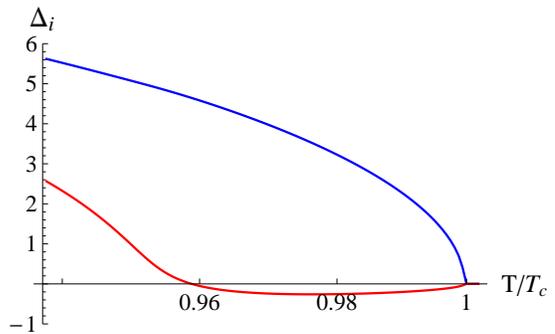}
\end{array}$
\end{center}
\caption{The behavior of $\Delta_1$ (blue) and $\Delta_2$ (red) with temperature,
demonstrating the $s_{\pm}$-$s_{\mathrm{++}}$ transition; $\Delta_2$ is negative
close to $T_c$, but goes through zero and changes its sign. The coupling constants
are $\lambda_{11}=0.3$, $\lambda_{22}=0.297$, $\lambda_{12}=-0.011$, $\lambda_{21}=-0.011$,
and $\Gamma=1.63$ (at the $T_{\gamma}$ $\Gamma\approx 1.67$).}
\label{Fig3}
\end{figure}

Below the critical line the quartic terms in the theory
become important. Let us consider a system with $T_c$ slightly higher than
$T_{\gamma}$ (meaning that immediately below $T_c$ it is in the $s_{\pm}$ state). If
$a_{22}(T)$ is positive then $\Delta_2$ is non-zero solely because of its coupling
to $\Delta_1$ through $a_{12}$. In the vicinity of $T_{\gamma}$ we can keep only the
linear in $\Delta_2$ terms in the equation $\partial
\mathcal{F}_{GL}/\partial|\Delta_2|=0$ (while keeping the cubic in $\Delta_1$
terms), and at the $s_{\pm}$ side we get:
\begin{eqnarray}
|\Delta_2|=-\frac{a_{12} + c_{11}|\Delta_1|^2}{ a_{22} + c_{12}|\Delta_1|^2 + b_{12}|\Delta_1|^2}|\Delta_1|.
\label{fi1}
\end{eqnarray}
It is clear that equation $a_{12} + c_{11}|\Delta_1|^2=0$ defines a line in the
$(\Gamma, T)$ space, originating from  $T_{\gamma}$, and separating the $s_{\pm}$
from the $s_{\mathrm{++}}$ regions. On this line the bands are decoupled and
$\Delta_2$ is zero.  If, for a fixed $\Gamma$, given system has $T_c$ slightly
higher than $T_{\gamma}$, with decreasing the temperature it will cross the line,
and $\Delta_2$ will change its sign. We demonstrate this in Fig. \ref{Fig3}. At this
$s_{\pm}$-$s_{\mathrm{++}}$ transition point the second band becomes normal again
(remember that we are assuming that $a_{22}(T)$ is still positive). Note however,
that neither of the gap parameters have any singularity at this point; in
thermodynamic sense this is a crossover, rather than a real phase transition.

What happens if, with decreasing the temperature, the system gets close to the
$a_{22}(T)=0$ point \emph{before} the $s_{\pm}$-$s_{\mathrm{++}}$ transition occurs? It can be easily shown that on the  $a_{12} + c_{11}|\Delta_1|^2=0$ line the $|\Delta_2|=0$ solution becomes unstable, and non-zero and purely imaginary $\Delta_2$ appears when $a_{22} - c_{12}|\Delta_1|^2 + b_{12}|\Delta_1|^2$ turns negative.
Since $\Delta_2$ is now a superconducting gap in
its own right, we have to keep all cubic terms in the equations. More generally, apart from the always-present $0$ and $\pi$ solutions, $\varphi$ can now take nontrivial values. From the condition $\partial\mathcal{F}_{GL}/\partial\varphi=0$ we obtain for $\varphi$ the equation:
\begin{eqnarray}
\cos\varphi=-\frac{a_{12} + c_{11}|\Delta_1|^2 + c_{22}|\Delta_2|^2}{2 c_{12}|\Delta_1||\Delta_2|}.
\label{cosfi}
\end{eqnarray}
This solution represents a distinct, intrinsically complex superconducting state.
The physical picture behind it is simple; instead of changing  the relative sign of
the gaps by taking one of them through zero, there is alternative, more elegant way
-- continuous evolution of $\varphi$ from $\pi$ to $0$. This intermediate
superconducting state can be understood as a linear combination (with complex
coefficients) of the two ``real" order parameters $s_{\pm}$ and $s_{\mathrm{++}}$.
More physically, this means that the fluctuations in the densities of the two
condensates (which are induced by fixing the phases) are not in-phase, as in
$s_{\mathrm{++}}$, and not in anti-phase, as in the $s_{\pm}$, but have some
nontrivial time shift. One of the modes is lagging the other, and as a consequence
the time-reversal symmetry is spontaneously broken (as it should in such
intrinsically complex state).  It is also easy to understand why such state appears
at finite temperature below $T_c$; close to the critical line only the $s_{\pm}$
state exists.  For the $s_{\mathrm{++}}$ state to condense within the $s_{\pm}$
state $a_{22}(T)$ has to turn negative, and only then the complex admixture of
$s_{\pm}$ and $s_{\mathrm{++}}$ becomes possible. This strongly suggests the
necessary condition for the existence of such complex state -- the presence of
\emph{two} attractive superconducting channels at the same temperature (which means
that $w$ has to be positive).
\begin{figure}[h]
\begin{center}$
\begin{array}{cc}
\includegraphics[width=0.4\textwidth]{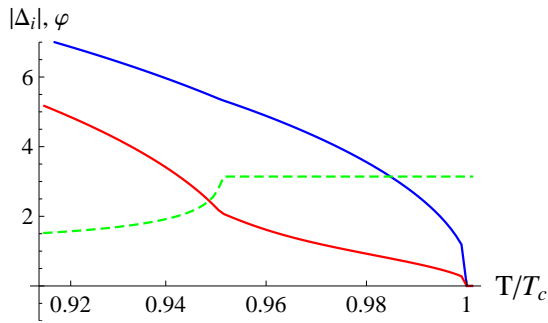}
\end{array}$
\end{center}
\caption{The behavior of $|\Delta_1|$ (blue), $|\Delta_2|$ (red) and $\varphi$
(green, dashed), for the same $\hat{\lambda}$ as in Fig. \ref{Fig3}, but for
$\Gamma=1.57$. Close to $T_c$ the relative phase is $\pi$ (the system is in the
$s_{\pm}$-state), but around $0.95 T_c$ it starts decreasing continuously. Both gaps
stay finite.} \label{Fig4}
\end{figure}

By minimizing the GL free energy, we demonstrate that this solution is indeed
realized, as illustrated in Fig. \ref{Fig4}. The order parameter starts as $s_{\pm}$
($\varphi=\pi$) at the critical temperature. However, at some finite temperature
below $T_c$ $\varphi$ deviates from the $\pi$ solution, and superconducting state is
no longer pure $s_{\pm}$, but an intrinsically complex state. According to our
model, the time-reversal symmetry breaking state is separated from the both ``real''
order parameters (which preserve the symmetry) by lines of continuous phase
transitions.

Similar complex states in one-band systems ($s+ id$ states)\cite{sid1, sid2, sid3,
sid4} and in three-band  systems ($s+is$ states) \cite{Agterberg, VS, Tanaka2, Hu,
Dias, Orlova, Maiti, Babaev2} have attracted recently a lot of attention. There are
some similarities in the underlying physics between these states and $s_{\pm} +
is_{\mathrm{++}}$ state discussed here. As in the $s+id$ case, in our model the
complex state appears as a way of avoiding the appearance of non-superconducting
parts of the Fermi surface (either the nodes of the $d$-wave state, or an entire
band in our model). The similarity with the three-band model is that in both cases
the complex order parameter admixes two superconducting states in the trivial
$A_{1g}$ representation. Our impurity-induced complex state is also somewhat similar
to the surface complex state predicted in the case of strong interband reflection at
the boundary\cite{Bobkovi}.

We summarize our findings in the phase diagram presented in Fig. \ref{Fig1}.
Strictly speaking, our results are valid only in the region of applicability of the
extended GL theory. To observe the complex state in this region we had to keep
$\lambda_{11}$ and $\lambda_{22}$ quite close. %
In the case they are not close the complex state is realized at temperatures
significantly lower than $T_c$ and has to be treated within the full microscopic
theory.
Nevertheless, using analogy with the physics and the phase diagrams discussed in
Refs. \onlinecite{VS, Maiti} we make two conjectures: i) the $s_{\pm} +
is_{\mathrm{++}}$
state is present if the system has $s_{\pm}$ to $s_{\mathrm{++}}$ crossover, %
even if it's not observable in the GL region; ii) this state extends down to $T=0$,
without any significant modifications. Confirming or rejecting these conjecture is
an important direction for future work.

What do our results imply for the iron-based
superconductors? Recently a roughly universal complete suppression of $T_c$ was
reported for several FeAs-122 compounds\cite{Kirshenbaum}.  This suggests that these materials are
in the $s_{\pm}$ state with strong interband pairing, and thus no complex state is
expected there. On the other hand, substantial variations in the effects of different impurities in
similar 122 systems were observed in Ref.\
\onlinecite{Li}. Also a very recent study of $T_c$ suppression in iron
chalcogenides\cite{Inabe} showed a non-universal behavior; with some of the
curves showing $T_c$ which initially decreases, but eventually saturates, as
expected for the $s_{\pm}$ to $s_{\mathrm{++}}$ transition. Although more studies
are needed, it is already clear that these materials are surprisingly diverse in their
normal and superconducting state properties, so it is entirely possible that the
$s_{\pm} + i s_{\mathrm{++}}$ state can be induced by impurities (for example, by
systematically irradiating a sample) in some of them.

In conclusion, we studied the role of impurities in a two-band superconductor. We
derived a Ginzburg-Landau theory to describe the system, and we showed that the
interband impurity scattering has a significant impact on the theory. Due to the
impurities-induced $\cos2 \varphi$ term in the theory a complex order parameter may
appear between the $s_{\pm}$ and $s_{\mathrm{++}}$ states.

This work was supported by by UChicago Argonne, LLC,
operator of Argonne National Laboratory, a U.S. Department of Energy Office of
Science laboratory, operated under contract No. DE-AC02-06CH11357, and by the Center
for Emergent Superconductivity, a DOE Energy Frontier Research Center, Grant No.
DE-AC0298CH1088.

\appendix

\section{Supplemental Material}
We start our derivation of the GL free energy from the Usadel equations for
the quasiclassical Greens functions $f(\mathbf{k},\mathbf{r},\omega)$ and
$g(\mathbf{k},\mathbf{r},\omega)$ \cite{Gurevich}. We only study uniform
states so these functions reduce to $f(\omega)$ and $g(\omega)$. In the
two-band case the equations have the form:
\begin{equation}
\omega f_{m}=\Delta_{m}g_{m}+\gamma_{mn}(g_{m}f_{n}-g_{n}f_{m}),
\label{UE1}
\end{equation}
where $m,n=(1,2)$ are the band indices and $m\neq n$ is implied. Notice that we are
treating the impurities in the Born approximation. We do not expect going beyond
that approximation to qualitatively change our result.

These equations have to
to supplemented by the self-consistency equations for the gap parameters
$\Delta_{1}$ and $\Delta_{2}$:
\begin{equation}
\Delta_{m}=2\pi T\sum_{n}\sum_{\omega>0}^{\omega_{0}}\lambda_{mn}f_{n},
\label{UE3}
\end{equation}
and normalization condition
\begin{equation}
|f_{m}|^{2}+g_{m}^{2}=1.
\end{equation}
To derive the GL equations we solve Eqs. \eqref{UE1} for $f_{1}$ and $f_{2}$,
and expand the solutions in powers of $\Delta_{1}$ and $\Delta_{2}$. To do
this we also have to expand $g_{m}$'s:
\[
g_{m}=\sqrt{1-|f_{m}|^{2}}\approx1-\frac{|f_{m}^{(0)}|^{2}}{2}%
\]
where $f_{m}^{(0)}$ is the zero-th order approximation:
\begin{equation}
f_{m}^{(0)}=\frac{(\omega+\gamma_{nm})\Delta_{m}+\gamma_{mn}\Delta_{n}}%
{\omega(\omega+\gamma_{mn}+\gamma_{nm})}. \label{UE5}
\end{equation}
Next order corrections are unwieldy, but straightforward to obtain. For
$f_{m}^{(1)}$ we get:
\begin{widetext}
\begin{eqnarray}
f^{(1)}_{m}=\frac{\gamma_{mn}(\omega+\gamma_{nm})(\Delta_m-\Delta_n)|f^0_{n}|^2
-\left[((\omega+\gamma_{nm})^2+ \gamma_{mn}(\omega + 2 \gamma_{nm}))\Delta_m
+ \gamma_{mn}(\omega+\gamma_{mn})\Delta_n \right]
|f^0_{m}|^2}{\omega(\omega+\gamma_{mn} + \gamma_{nm})}.
\end{eqnarray}
\end{widetext}
Inserting $f^{(0)}$, we get an expression for $f^{(1)}$ which is
of order $\Delta^{3}$. If we define
\[
R_{m}=2\pi T\sum_{\omega>0}^{\omega_{0}}(f_{m}^{(0)}+f_{m}^{(1)}),
\]
the self-consistency equations give:
\[
R_{m}=\frac{1}{w}(\lambda_{mm}\Delta_{m}-\lambda_{mn}\Delta_{n}),
\]
with $\det[\hat{\lambda}]\equiv w=\lambda_{11}\lambda_{22}-\lambda_{12}%
\lambda_{21}$. Expressing $R_{m}$ via $\Delta_{m}$ and $\Delta_{n}$, we get
two equation for the two gap parameters up to $\Delta^{3}$. They are identical
to the equations $\delta\mathcal{F}_{GL}/\delta\Delta_{m}^{\ast}=0$ obtained
by varying the GL free energy with respect to $\Delta_{m}^{\ast}$. Collecting
all the terms, multiplying by the density of states $N_{m}$, and using the
notation introduced in the main text, we get:
\begin{align}
&  a_{mm}\Delta_{m}+a_{mn}\Delta_{n}+b_{mm}\Delta_{m}|\Delta
_{m}|^{2}\nonumber\\
&  +b_{mn}\Delta_{m}|\Delta_{n}|^{2}+c_{mm}(\Delta_{m}^{2}\Delta
_{n}^{\ast}+2|\Delta_{m}|^{2}\Delta_{n})\nonumber\\
&  +c_{nn}|\Delta_{n}|^{2}\Delta_{n}+c_{mn}\Delta_{m}^{\ast}%
\Delta_{n}^{2}=0.
\end{align}
The coefficients are defined as follows:
\[
a_{mm}=N_{m}\left(  \frac{\lambda_{nn}}{w}-2\pi T\sum_{\omega>0}^{\omega_{0}%
}\frac{\omega+\gamma_{mn}}{\omega(\omega+\gamma_{mn}+\gamma_{nm})}\right)  ,
\]%
\[
a_{mn}=-N_{m}\left(  \frac{\lambda_{mn}}{w}+2\pi T\sum_{\omega>0}^{\omega
_{0}}\frac{\gamma_{mn}}{\omega(\omega+\gamma_{mn}+\gamma_{nm})}\right)  ,
\]
\begin{widetext}
\begin{equation}
b_{mm}=N_{m}\pi T\sum_{\omega>0}^{\omega_{0}}\frac{(\omega+\gamma_{nm})^{4}%
}{\omega^{3}(\omega+\gamma_{mn}+\gamma_{nm})^{4}}+N_{m}\pi T\sum_{\omega
>0}\frac{(\omega+\gamma_{nm})\gamma_{mn}\left(  \omega^{2}+3\omega\gamma
_{nm}+\gamma_{nm}^{2}\right)  }{\omega^{3}(\omega+\gamma_{mn}+\gamma_{nm}%
)^{4}},\nonumber
\end{equation}%
\begin{equation}
b_{mn}=-N_{m}\pi T\sum_{\omega>0}^{\omega_{0}}\frac{\gamma_{mn}\omega^{3}%
}{\omega^{3}(\omega+\gamma_{mn}+\gamma_{nm})^{4}}+\ N_{m}\pi T\sum_{\omega
>0}\frac{\gamma_{mn}(\gamma_{mn}+\gamma_{nm})(\gamma_{nm}(\omega+2\gamma
_{mn})+\omega\gamma_{mn}))}{\omega^{3}(\omega+\gamma_{mn}+\gamma_{nm})^{4}%
},\nonumber
\end{equation}%
\[
c_{mm}=N_{m}\pi T\sum_{\omega>0}^{\omega_{0}}\frac{\gamma_{mn}(\omega
+\gamma_{nm})\left(  \omega^{2}+(\omega+\gamma_{nm})(\gamma_{mn}+\gamma
_{nm})\right)  }{\omega^{3}(\omega+\gamma_{mn}+\gamma_{nm})^{4}},
\]
\end{widetext}
\[
c_{mn}=N_{m}\pi T\sum_{\omega>0}^{\omega_{0}}\frac{\gamma_{mn}(\omega
+\gamma_{mn})(\omega+\gamma_{nm})(\gamma_{mn}+\gamma_{nm})}{\omega^{3}%
(\omega+\gamma_{mn}+\gamma_{nm})^{4}}%
\]
The sums for all coefficients can be carried out, and closed-form analytic
results can be obtained. Unfortunately, these results are complicated
combinations of polygamma functions (digamma function and its derivatives),
and since they do not provide any further insight into the problem, we will
not show them.

For a fixed coupling constants matrix $\hat{\lambda}$ and disorder strength
$\Gamma$ all coefficients are functions of temperature. The sign change of
$a_{11}$ and $a_{22}$ drives the superconducting transition, and the sign
change of $a_{12}$ drives the $s_{\pm}$-to-$s_{++}$ crossover. Close to
$T_{c}$ the quartic coefficients are only weakly temperature dependent, and,
with the exception of $b_{12}$, are all positive. In addition, $c_{12}$ tends
to be the smallest.

As emphasized in the main text, in the limit $\Gamma\rightarrow0$ all quartic
coefficients that couple $\Delta_{1}$ and $\Delta_{2}$ vanish, and we recover
the clean two-band GL theory. For non-zero $\Gamma$, however, we have to use
the full free energy $\mathcal{F}_{GL}$.

Close to $T_{c}$ only the linear terms matter. From Eqs. \eqref{UE3} and
\eqref{UE5} we obtain the self-consistency equations for the two-band case
\[
\Delta_{m}=2\pi T\sum_{n}\sum_{\omega>0}^{\omega_{0}}\lambda_{mn}\frac
{(\omega+\gamma_{\bar{n}n})\Delta_{n}+\gamma_{n\bar{n}}\Delta_{\bar{n}}%
}{\omega(\omega+\gamma_{n\bar{n}}+\gamma_{\bar{n}n})}%
\]
with $\bar{n}=1(2)$ for $n=2(1)$. These equations can be represented in the
form of the matrix equation used in the main text,
\begin{equation}
\Delta_{m}=\sum_{n}\mathbb{M}_{mn}\Delta_{n} \label{Tc1}%
\end{equation}
where the matrix $\mathbb{M}_{mn}$ is given by
\[
\mathbb{M}_{mn}=\lambda_{mn}I_{2}+\lambda_{m}n_{n}I_{-}.
\]
Here we have used the relation $n_{m}=\gamma_{nm}/(\gamma_{mn}+\gamma_{nm})$,
and defined $I_{-}=I_{1}-I_{2}$, with
\[
I_{1}=2\pi T\sum_{0}^{\omega_{0}}\frac{1}{\omega_{n}},\ \ I_{2}=2\pi T\sum
_{0}^{\omega_{0}}\frac{1}{\omega_{n}+\gamma_{12}+\gamma_{21}}.
\]
The quantity $I_{-}$ can be expresses via the digamma function $\psi(x)$ as
$I_{-}=\psi\left(  \frac{1}{2}+\frac{N\Gamma}{2\pi T}\right)  -\psi(1/2)$ with
$N\Gamma=\gamma_{12}+\gamma_{21}$.

Eq.\ \eqref{Tc1} can also be used to derive an analytic formula for $T_{c}$ in
the extreme dirty limit. We rewrite this equation in somewhat different form,
more convenient for analytical analysis. The sum $I_{1}$ can be represented as
$I_{1}=\ln(T_{c0}/T_{c})+1/\lambda$, where $\lambda=\frac{\lambda_{11}%
+\lambda_{22}}{2}+\sqrt{\frac{\left(  \lambda_{11}-\lambda_{22}\right)  ^{2}%
}{4}+\lambda_{12}\lambda_{21}}$ is the largest eigenvalue of $\hat{\lambda}$,
which determines the clean-limit transition temperature, $T_{c0}$. This allows
us to represent the matrix $\mathbb{M}$ as
\[
\mathbb{M}_{mn}=\lambda_{mn}\left(  \ln\frac{T_{c0}}{T_{c}}+\frac{1}{\lambda
}\right)  -\lambda_{mn}I_{-}+\lambda_{m}n_{n}I_{-}.
\]
Multiplying both sides of the matrix equation (\ref{Tc1}) with $\hat{\lambda
}^{-1}$ and using $\lambda_{mn}^{-1}\lambda_{n}=1$, we obtain
\[
\sum_{n}\lambda_{mn}^{-1}\Delta_{n}=\left(  \ln\frac{T_{c0}}{T_{c}}+\frac
{1}{\lambda}-I_{-}\right)  \Delta_{m}+I_{-}\sum_{n}n_{n}\Delta_{n}.
\]
Introducing notation $w_{mn}=\lambda_{mn}^{-1}-\lambda^{-1}\delta_{mn}$, where
$\delta_{mn}$ is the Kronecker delta, we can cast this in an equivalent form:
\begin{equation}
\sum_{n}\left(  w_{mn}\!-\ln\frac{T_{c0}}{T_{c}}\delta_{mn}\right)\!  \Delta
_{n}\!=\!-I_{-}\sum_{n}n_{n}\left(  \Delta_{m}\!-\!\Delta_{n}\right)
.\label{DeltaSyst}%
\end{equation}
General equation for $T_{c}$ is determined by vanishing of the determinant for
this linear system which gives%
\begin{align}
&  \ln\frac{T_{c0}}{T_{c}}\left(  w_{11}+w_{22}+I_{-}-\ln\frac{T_{c0}}{T_{c}%
}\right)  \nonumber\\
&  =I_{-}\left[  n_{1}\left(  w_{11}+w_{12}\right)  +n_{2}\left(
w_{22}+w_{21}\right)  \right].  \label{EqTc}%
\end{align}

In the dirty limit, $N\Gamma \gg T_c$, we can use the asympotics of $I_{-}$,
$I_{-}\approx\ln \frac{{N\Gamma}}{AT_{c}}$ with $A=\pi\exp(-\gamma_{E})/2$. In this
case
we obtain from Eq. (\ref{EqTc})%
\[
\ln\frac{T_{c0}}{T_{c}}=\frac{\left[  n_{1}\left(  w_{11}+w_{12}\right)
+n_{2}\left(  w_{22}+w_{21}\right)  \right]  \ln\frac{{N\Gamma}}{AT_{c0}}%
}{n_{2}\left(  w_{11}-w_{21}\right)  +n_{1}\left(  w_{22}-w_{12}\right)
+\ln\frac{{N\Gamma}}{AT_{c0}}}.
\]
In the extreme dirty case corresponding to condition $\ln\frac{{N\Gamma
}}{AT_{c0}}\gg n_{2}\left(  w_{11}-w_{21}\right)  +n_{1}\left(  w_{22}%
-w_{12}\right)  $, we obtain for the limiting value of transition temperature,
$T_{c\infty}$,%
\[
\ln\frac{T_{c0}}{T_{c\infty}}=n_{1}\left(  w_{11}+w_{12}\right)  +n_{2}\left(
w_{22}+w_{21}\right)
\]
This result actually can be obtained directly from Eq.\ (\ref{DeltaSyst}) if we take
$\Delta_{1}\!=\!\Delta_{2}$ in the left-hand side. Using the definition of $T_{c0}$,
the above result for $T_{c\infty}$ can be rewritten in
somewhat more transparent form%
\[
\ln\frac{\omega_{0}}{AT_{c\infty}}=\frac{n_{1}\left(  \lambda_{22}%
-\lambda_{12}\right)  +n_{2}\left(  \lambda_{11}-\lambda_{21}\right)  }{w}.
\]
The quantity in the right-hand side represents the band-average of the inverse
coupling constant $\left\langle \lambda^{-1}\right\rangle $.

Now let us derive the formula for $T_{\gamma}$ shown in the main text.
Remember that $T_{\gamma}$ is defined as the $T_{c}$ point at which $a_{12}$
coefficient vanishes. This happenes at:
\begin{align}
w_{12} = - \frac{\lambda_{12}}{w} = n_{2} I_{-}.\nonumber
\end{align}
Using this condition in the general $T_{c}$ formula given above we get:
\begin{align}
\ln\frac{T_{c0}}{T_{\gamma}}=w_{11}+w_{12},\nonumber
\end{align}
and combining this with the clean limit expression $\ln( \omega_{0}/ A
T_{c0})=\lambda^{-1}$ gives the formula in the main text
\begin{align}
T_{\gamma}\approx1.13\omega_{0}\exp\left[  -\frac{\lambda_{22}-\lambda_{12}%
}{w}\right]  .
\end{align}

 \bibliographystyle{apsrev}

\begin{thebibliography}{10}
\bibitem{crossover} Strictly speaking, in a thermodynamic sense,  this is not a true phase transition, but rather a crossover.
\bibitem{GLnote}Typically, superconducting instability dominantly develops in one
channel and near $T_{c}$ corrections from other channels can be neglected. In this
case the two-band GL equations can be reduced to classical single-order parameter
description\cite{Kogan1}. In some range of parameters, however, superconducting
instabilities for two bands may occur at very close temperatures. In this case,
which we consider in this Letter, the singe-order-parameter description shrinks to
the very narrow range near $T_c$ and one has to use the full two-band GL equations.
\end{thebibliography}

\end{document}